\documentclass[aps,groupedaddress,superscriptaddress,amsmath,amssymb,twocolumn,pra]{revtex4-1}
\usepackage{bm}
\usepackage[dvips]{graphicx}
\usepackage{wrapfig,subfigure}
\usepackage{stmaryrd}
\usepackage{color}
\usepackage[normalem]{ulem}
\usepackage{enumerate}
\usepackage{epstopdf}
\def\qb{{\bf q}}

\def\rb{{\bf r}}
\def\pb{{\bf p}}
\def\qb{{\bf q}}
\def\vep{\varepsilon}

\newcommand{\Ket}[1]{\vert #1 \rangle}
\newcommand{\Bra}[1]{\langle #1 \vert}
\def\n{{\bm\nabla}}

\begin{document}

\title{Ripplonic Lamb shift for electrons on liquid helium}

\author{D. Konstantinov}
\affiliation{Quantum Dynamics Unit, Okinawa Institute of Science and Technology, Tancha 1919-1, Okinawa 904-0495, Japan}
\author{K.  Kono}
\affiliation{RIKEN CEMS - Hirosawa 2-1, Wako 351-0198, Japan}
\author{M. J. Lea}
\affiliation{Department of Physics, Royal Holloway, University of London, TW20 0EX, UK}
\author{M. I.  Dykman}
\affiliation{Department of Physics and Astronomy, Michigan State University, MI 48824, USA}
\date{\today}
\begin{abstract}
We study the shift of the energy levels of electrons on helium surface due to the coupling to the quantum field of surface vibrations. As in quantum electrodynamics, the coupling is known, and it is known to lead to an ultraviolet divergence of the level shifts. We show that there are diverging terms of different nature and use the Bethe-type approach to show that they cancel each other, to the leading-order. This  resolves the long-standing theoretical controversy and explains the existing experiments. The results allow us to study the temperature dependence of the level shift. The predictions are in good agreement with the experimental data.
\end{abstract}
\maketitle

Electrons above the surface of liquid helium were one of the first observed two-dimensional electron systems (2DESs) \cite{Williams1971,Sommer1971,Brown1972,Grimes1976a}. In this system the conceptual simplicity is combined with far from trivial behavior, which allows studying many-body effects in a well-characterized setting. The system displays the highest mobility known for 2DESs, exceeding $2\times 10^8$~cm$^2$/(V$\cdot$s) \cite{Shirahama1995} and can be exquisitely well controlled \cite{Bradbury2011}. The electron-electron interaction is typically strong, so that the electrons can form a Wigner solid \cite{Grimes1979,Fisher1979} or a strongly correlated liquid with unusual transport properties \cite{Dykman1979c,Dykman1993i}. A number of new many-electron phenomena have been found recently \cite{Konstantinov2009,Ikegami2012,Konstantinov2012a,Chepelianskii2015,Rees2016}.

An advantageous feature of the system is the simple form of the confining potential. It is formed by the high Pauli barrier at the helium surface and the image potential. One then expects the electron energy spectrum to be well understood. Indeed, already the first experiment on transitions between the subbands of quantized motion normal to the surface showed a good, albeit imperfect agreement with the model \cite{Grimes1976a}. Much work has been done on improving, sometimes empirically, the form of the confining potential, cf. \cite{Grimes1976a,Stern1978,Rama1988,Cheng1994,Degani2005}. On the other hand, it has been known that the electrons are also coupled to a bosonic field, the capillary waves on the surface of helium (ripplons), and that this coupling affects the electron energy spectrum \cite{Shikin1971,Shikin1974a,Jackson1981,Jackson1982,Klimin2016}. The importance of this effect was demonstrated in explaining the Wigner crystallization \cite{Fisher1979,Yuecel1993} and through cyclotron-resonance measurements \cite{Wilen1988a}.

In terms of the coupling to a bosonic field, electrons on helium are a close condensed-matter analog of systems studied in quantum electrodynamics. The known form of the coupling \cite{Andrei1997} and the possibility to control it and to study the interplay of this coupling with the many-electron effects make the system particularly attractive. In this context, a major problem is that the ripplon-induced shift of the electron energy levels of motion normal to the surface contains diverging terms. They come from short-wavelength ripplons. The ultraviolet divergence is strong, as a high power of the wave number. Unless one deals with it carefully, the resulting level shifts become comparable to the electron binding energy for a short-wavelength  cutoff approaching twice the interatomic distance. The problem bears also on 2DESs in other systems, where the barrier at the interface is high whereas the interface itself is randomly warped.

In this paper we show that, in fact, the level shift due to the coupling to ripplons is small. The situation is reminiscent of the Lamb shift in QED. We show that there are two groups of diverging terms and use the Bethe trick to demonstrate the compensation of the leading ultraviolet-divergent terms in the overall level shift.  Interestingly, the remaining correction still displays a power-law ultraviolet divergence, which is much weaker, however. This shows the nontrivial nature of the compensation. Once the account is taken of the natural cutoff at the interatomic distance, the correction becomes small. It describes a Lamb-shift type deviation from the energy spectrum in the absence of the electron-ripplon coupling.

 We also study the temperature-dependent shift of the energy levels. We find a good agreement with the experimental results on the spectra of inter-subband transitions \cite{Collin2002} and the new results \cite{Collin2017} that corroborate the theory. The theory also explains the long lifetime of the electron states, which are critical for a potential implementation of a quantum computer based on electrons on helium \cite{Platzman1999,Dykman2003,Schuster2010a}. The proposed divergence-compensation mechanism is fairly general for quasi-two-dimensional systems. 

The contributions to the level shift due to the warping of the helium surface, i.e., to the ripplons, can be separated into three parts. One is kinematic and comes from the electron kinetic energy over a warped surface. The other is electrostatic and comes from the difference of the electron potential energies above the plane and a warped surface. The third comes from the intertia of the surface waves. The key point of the analysis of these contributions is that we have to go beyond the standardly used polaron theory in which the electron system is assumed to be two-dimensional. It is necessary to take into account the ripplon-induced mixing of the different states of motion normal to the surface.

\begin{figure}[h]
\includegraphics[width=4.2cm]{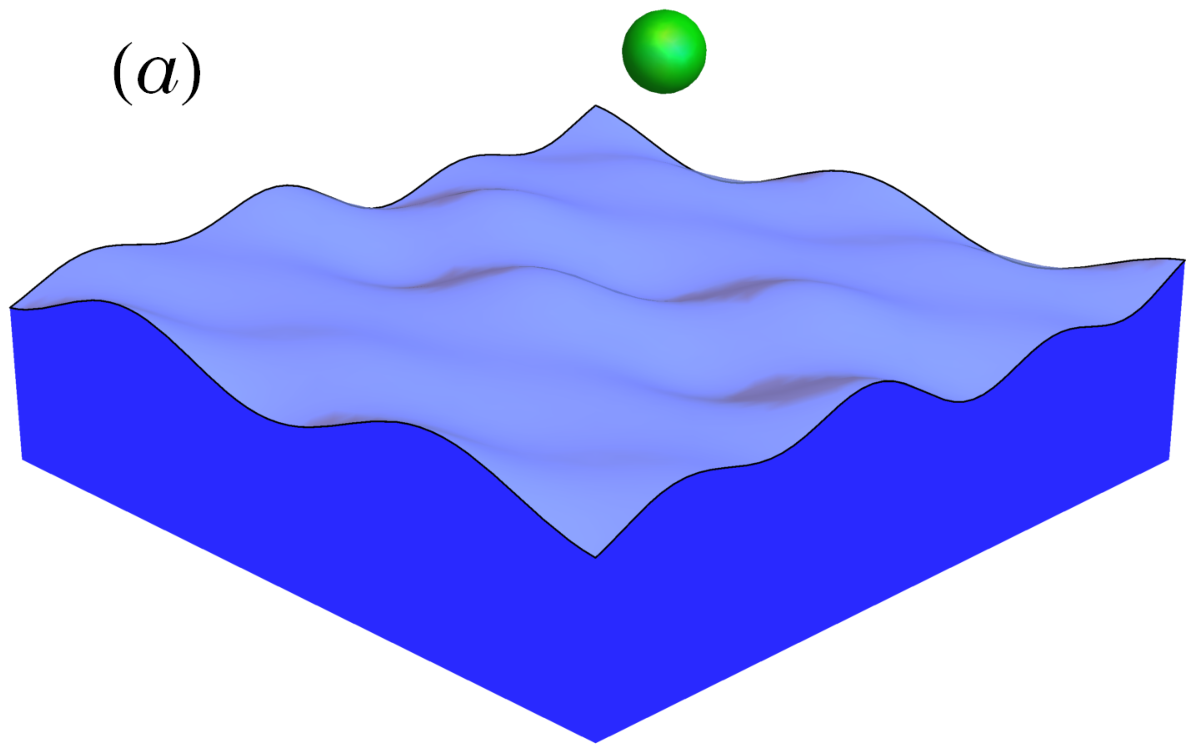}\hfill
\includegraphics[width=3.6cm]{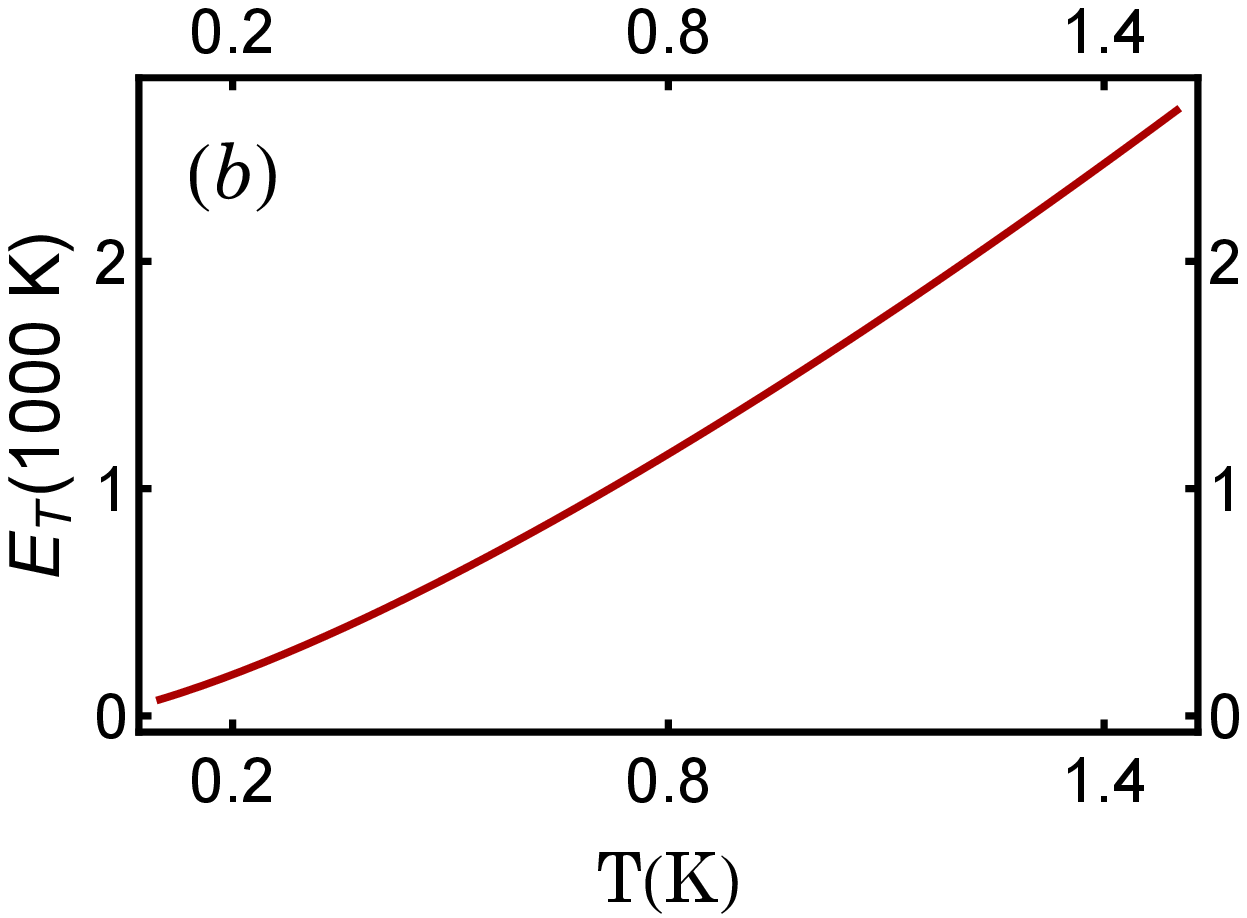}\\
\vspace*{0.1in}
\includegraphics[width=4.0cm]{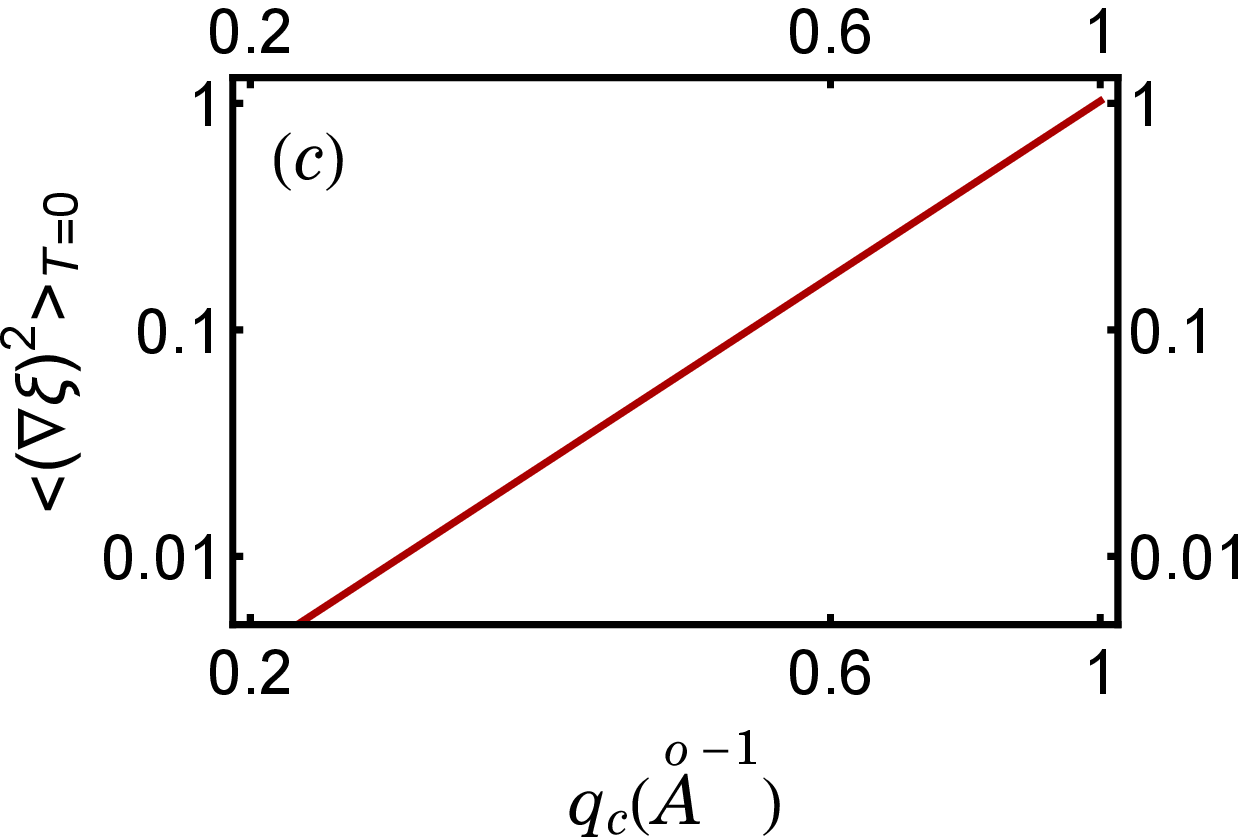}
\hfill\includegraphics[width=4.2cm]{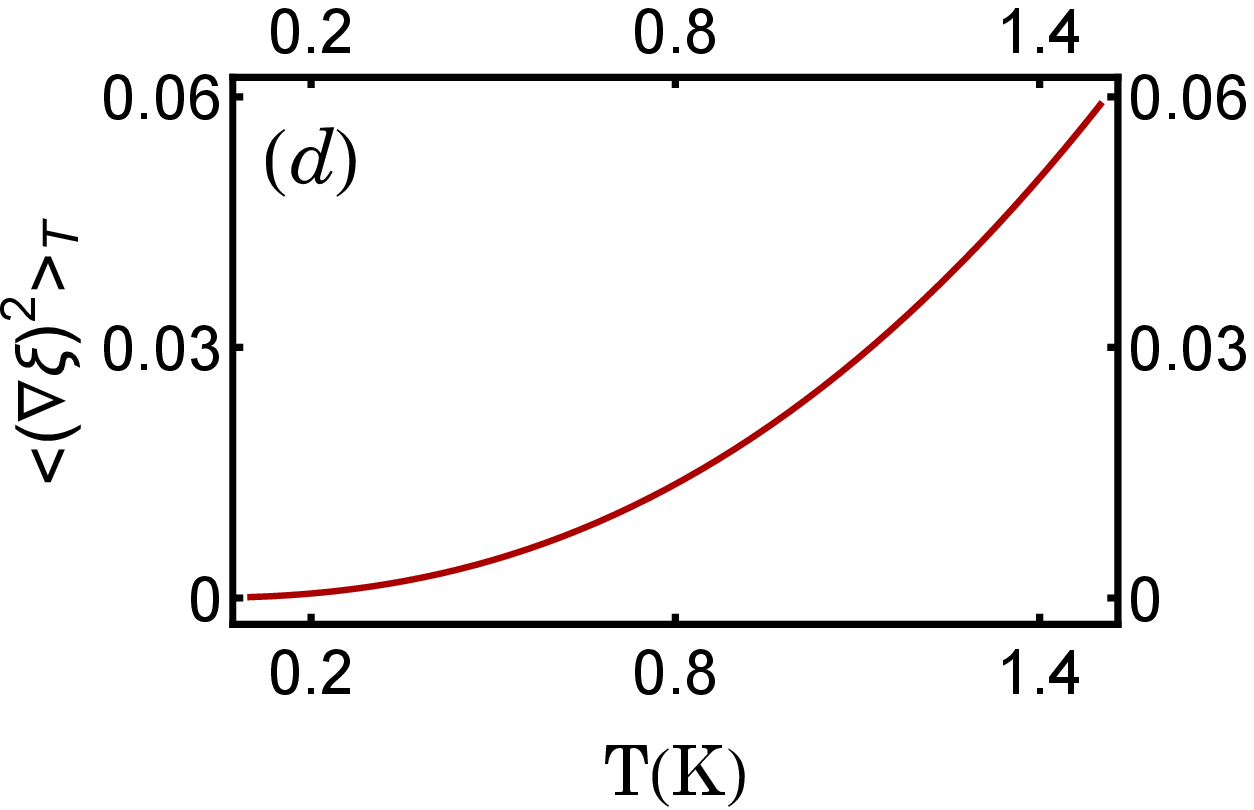}
\caption{(a) Sketch of an electron above warped helium surface. (b) Electron energy $E_{\rm T}=\hbar^2 q_T^2/2m$ for a thermal ripplon wave number $q_T$,  $\hbar \omega_{q_T}=k_BT$. (c) The leading term in the relative change of the reciprocal electron mass  due to the direct kinematic two-ripplon coupling $\langle h_i^{(2)}\rangle $, Eq.~(\ref{eq:correction_2}), for $T=0$. Plotted is $\langle (\n\xi)^2\rangle)_{T=0}$ as function of the short-wavelength cutoff $q_c$. (d) Same as in (c), but when only the thermal contribution is taken into account.}
\label{fig:sketch}
\end{figure}

The kinematic level shift is the major one. To see how it comes about, we choose $\rb=(x,y)$ and $\pb=(p_x,p_y)$ as the two-dimensional vectors of motion parallel to the surface; the coordinate $z$ is normal to the surface, the electron motion along $z$ is quantized, see Fig.~\ref{fig:sketch}. Qualitatively, one can think that the nonuniform surface displacement $\xi(\rb)$ changes the effective width of the potential well in the $z$-direction. This changes the kinetic energy of the confined motion. Since $\langle \xi(\rb)\rangle = 0$, the change is quadratic in $\xi$, and in fact, in $\n\xi$, because a uniform surface displacement does not change the energy. The energy change is positive, since the kinetic energy scales as the squared reciprocal confinement length. 

On the other hand, the motions parallel and normal to the surface are mixed by the warping. In the second order (again, quadratic in $\xi$), the mixing leads to shifts of the levels of motion normal to the surface, which are negative for low-lying levels, as expected from the standard perturbation theory. It turns out that, taken separately, both mechanisms display strong ultraviolet divergences  of the energy shift, which partly compensate each other. 

The warping-induced change of the electrostatic energy has linear and quadratic in $\xi(\rb)$ terms as well. The shift of the electron energy levels due to the quadratic term also displays an ultraviolet divergence, but it is weaker than for the kinematic mechanism. This is because the image potential is much less sensitive to short-wavelength fluctuations of the helium surface. The divergence is partly compensated by the single-ripplon processes, as for the kinematic coupling. The analysis is similar to the analysis of the kinematic effect given below and is provided in the Supplemental Material \footnote{The two-ripplon processes due to the electrostatic electron-ripplon coupling are discussed in the Supplemental Material (SM). The SM also provides the details of the numerical calculations of the relevant matrix  elements and the semiclassical approximation for highly excited states}.

Our starting point is the electron Hamiltonian for a flat helium surface,
\begin{equation}
\label{eq:Hamiltonian_flat}
H_0= \hat T +V(z),\qquad \hat T=(2m)^{-1}(\pb ^2 + p_z^2),
\end{equation}
where the potential energy  $V(z)$ for $z>0$ comes from the image force and from the electric field $E_\perp$ usually applied to press the electrons to the surface \cite{Andrei1997,Monarkha2004}. The potential has an atomically steep repulsive barrier at $z=0$ with height $> 1$~eV formed by helium atoms. The eigenstates of Hamiltonian (\ref{eq:Hamiltonian_flat}) are products of the wave functions of quantized motion normal to the surface $|n\rangle\equiv\psi_n(z)$ and the plane waves of lateral motion $\propto \exp(i\pb\rb/\hbar)$. The energies of the normal and lateral motions are $E_n$ and $E_\pb$, the total energy is $E_{n\pb}=E_n+E_\pb$.

The ripplon Hamiltonian and the ripplon-induced surface displacement are 
\begin{equation}
\label{eq:ripplon_Hamiltonian}
\hat H_r=\hbar\sum_\qb \omega_q  b^\dagger_\qb b_\qb, \qquad \xi(\rb)=\sum_\qb Q_qe^{i\qb\rb}(b_\qb + b_{-\qb}^\dagger).
\end{equation}
Here, $b_\qb$ is the annihilation operator of a ripplon with the wave number $\qb$ and frequency $\omega_q$; $Q_q= (\hbar q/2\rho\omega_qS)^{1/2}$, where $\rho$ is the helium density and $S$ is the area \cite{Cole1974}.

The effect of the ripplon-induced curvature of the electron barrier at the helium surface can be taken into account in a standard way \cite{Migdal2000} by making a canonical transformation $U=\exp[-i\xi(\rb)p_z/\hbar]$, which shifts the electron $z$-coordinate so that it is counted off from the local position of the surface, $z\to  z-\xi(\rb)$ \cite{Shikin1974a}. 
The transformed electron kinetic energy and the ripplon energy $U^\dagger (\hat T +\hat H_r)U$ is the sum $\hat T+ \hat H_r + \hat H_i^{(1)} + \hat H_i^{(2)}$, where $\hat H_i^{(1)}$ and $\hat H_i^{(2)}$ describe the linear and quadratic in $\xi(\rb)$ kinematic electron-ripplon coupling, 
\begin{align}
\label{eq:transformed_kinetic}
&\hat H_i^{(1)}=-\frac{1}{2m}p_z \{\pb,\n\xi(\rb)\}_+ + i\hbar^{-1}p_z[\xi(\rb),H_r]  ,\nonumber\\
&\hat H_i^{(2)}= p_z^2h_i^{(2)}, \quad h_i^{(2)}=\frac{1}{2m}(\n\xi)^2 + \sum_\qb\omega_q|Q_q|^2/\hbar
\end{align} 
[$\{A,B\}_+ = AB+BA, \n = (\partial_x,\partial_y)$ ]. 

The term $\hat H_i^{(2)}$ averaged over the thermal distribution of ripplons gives a correction to the electron energy $E_n$ already in the first order,
\begin{equation}
\label{eq:correction_2}
\Delta E_n^{(2)}=\Bra{n}p_z^2\Ket{n}\langle h_i^{(2)}\rangle.
\end{equation}
This correction is just a renormalization of the electron mass for motion normal to the surface $m^{-1}\to m^{-1}+ \sum_\qb |Q_q|^2\left[2m\omega_q + \hbar q^2 (2\bar n_q+1)\right]/m\hbar$, where $\bar n_q=[\exp(\hbar\omega_q/k_BT)-1]^{-1}$ is the Planck number. 
Since $\omega_q\propto q^{3/2}$ for large $q$ \cite{LL_Hydrodynamics}, the sum over $\qb$  diverges as $q_c^{7/2}$ for $T=0$, where $q_c$ is the short-wavelength cutoff. As seen from Fig.~\ref{fig:sketch}, if we set $q_c= 1\AA^{-1}$\footnote{Such cutoff corresponds to the estimate (K. R. Atkins, Can. J. Phys. {\bf 31}, 1165 (1953)) that $q_c^2/4\pi$ is the number of helium atoms in the surface monolayer. Ripplons with $q\lesssim 1.5~\AA^{-1}$ were seen in the neutron scattering spectra by  H. J. Lauter {\it et al.}, Phys. Rev. Lett. {\bf 68}, 2484 (1992).\label{fn:Atkins}}, the energy shift (\ref{eq:correction_2}) is $\Delta E_n^{(2)}\sim E_n$, indicating the breakdown of the perturbation theory. 

Operator $\hat H_i^{(1)}$ contributes to the level shift in the second order. For a state with given quantum numbers $n$ and $\pb$, the shift is 
\begin{align}
\label{eq:correction_1}
\Delta E_{n\pb}^{(1)}=&\sum_{n',\qb}\,\sum_{\alpha=\pm 1}{\cal N}_{q\alpha}\frac{|\Bra{n}p_z\Ket{n'}|^2 \Delta_{\pb,\qb,\alpha}^2}{E_n-E_{n'} - \Delta_{\pb,\qb,\alpha}}.
\end{align}
Here, $\alpha$ allows for the processes with virtual emission ($\alpha=1$) or absorption ($\alpha=-1$) of a ripplon, $\Delta_{\pb,\qb,\alpha}=E_{\pb +\hbar\qb} - E_\pb  + \alpha\hbar\omega_q$ is the difference of the energies of the in-plane motion in the initial and  intermediate electron states with the added or subtracted ripplon energy, and ${\cal N}_{q\alpha}= |Q_q|^2\left[\bar n_q+ (1+\alpha)/2\right]/\hbar^2$.

We find the overall kinematic energy shift $\Delta E_{n\pb}^{\rm kin}=\Delta E_{n\pb}^{(1)} + \Delta E_n^{(2)}$ by re-writing in Eq.~(\ref{eq:correction_1})
\begin{equation}
\label{eq:Bethe}
\frac{\Delta_{\pb,\qb,\alpha}}{E_n-E_{n'}-\Delta_{\pb,\qb,\alpha}}=\frac{E_n-E_{n'}}{E_n-E_{n'}-\Delta_{\pb,\qb,\alpha}}-1. 
\end{equation}
The last term here exactly cancels $\Delta E_n^{(2)}$ in $\Delta E_{n\pb}^{\rm kin}$, since $\sum_{n'}|\Bra{n}p_z\Ket{n'}|^2=\Bra{n}p_z^2\Ket{n}$ and $Q_q,\omega_q$ are independent of the direction of $\qb$. Then 
\begin{equation}
\label{eq:kinematic}
\Delta E_{n\pb}^{\rm kin}= \sum_{n',\qb,\alpha}{\cal N}_{q\alpha} \frac{\Delta_{\pb,\qb,\alpha}|\Bra{n}p_z\Ket{n'}|^2 (E_n-E_{n'})}{E_n-E_{n'} - \Delta_{\pb,\qb,\alpha}}.
\end{equation}
We will be interested in $\Delta E_{n\pb}^{\rm kin}$ for low-lying out-of-plane states, $n\sim 1$, and for small momenta $p\lesssim (mk_BT )^{1/2}$.

The unperturbed energies $E_n,E_{n'}$  and the matrix elements $\Bra{n'} p_z\Ket{n}$ in Eq.~(\ref{eq:kinematic}) can be found from the one-dimensional Schr\"odinger equation for an electron above the flat helium surface. To get an analytic insight we note that the main contribution to Eq.~(\ref{eq:kinematic}) comes from large in-plane wave numbers $q$ compared to the reciprocal out-of-plane localization length $r_B^{-1}$ in the ground state $n=1$ ($r_B\lesssim 100\AA$ \cite{Cole1970,Cole1974}).  The energy $\Delta_{\pb,\qb,\alpha}$ largely exceeds $k_BT$ and  $|E_n|$ with $n\sim 1$. Then of primary importance is the contribution to Eq.~(\ref{eq:kinematic})  of highly excited states with $n'\gg 1$. Such states are semiclassical. They correspond to electron motion in an  almost triangular potential well formed by the barrier at $z=0$ and the field $eE_{\perp}$ that presses the electrons against the surface. The WKB approximation gives $E_{n'} \approx [3\pi\hbar e E_{\perp}(n'-1/4)/2\sqrt{2m}]^{2/3}$ for $n'\gg 1$ (a better approximation is based on matching the WKB and the small-$z$ wave functions [33]). 
Since the large-$n'$ wave functions are fast oscillating on length $r_B$, one can show that $|\Bra{n'}p_z\Ket{n}|^2 \approx (\hbar^4 eE_{\perp}/2m E_{n'}^2)|\partial_z\psi_{n}|_{z=0}^2$. 

Changing from summation over $n'$ in Eq.~(\ref{eq:kinematic}) to integration, we obtain  from the above estimate
\begin{align}
\label{eq:asymptotic}
\Delta E_{n\pb}^{\rm kin}\approx \frac{\hbar }{\sqrt{2m}} |\partial_z\psi_{n}|_{z=0}^2 \sum_\qb|Q_q|^2(2\bar n_q+1)\Delta_{\pb,\qb,1}^{1/2}.
\end{align}
We disregarded here the contribution of the states with energies $E_{n'}\lesssim E_n$. We also disregarded the ripplon energy $\hbar\omega_q$ compared to the in-plane electron energy $E_{\hbar\qb}$. This is a good approximation, because ripplons are slow,   $\omega_q \approx (\sigma/\rho)^{1/2}q^{3/2}$, where $\sigma $ is the helium surface tension. Therefore their thermal momentum $q_T$ given by condition $\hbar\omega_{q_T}=k_BT$ corresponds to the electron energy $E_{\hbar q_T} $ varying from $\approx 73$~K to $\approx  1.6\times 10^3$~K for $T$ varying from 0.1~K to 1~K, see Fig.~\ref{fig:sketch}(b). 

The $T=0$-term in Eq.~(\ref{eq:asymptotic}) still has an ultraviolet divergence. It scales with the short-wavelength cutoff $q_c$ as $q_c^{5/2}$. This is a much weaker divergence than that of $\Delta E_n^{(2)}$. Overall, the $T=0$ term is  $ \sim \Delta E_n^{(2)}/(q_c r_B)$. For the cutoff $q_c\sim 1~\AA^{-1}$,  this term is on the order of a few percent of the electron binding energy $|E_1|=\hbar^2/2mr_B^2\sim 8$~K. Its dependence on the control parameter $E_\perp$ is described by Eq.~(\ref{eq:kinematic}). There are also other contributions to the $T=0$-level shift. They include the effect of the electron correlations \cite{Lambert1980,Konstantinov2009},  the intraband polaronic shift \cite{Jackson1981} and, last but not least, the finite steepness and height of the barrier for electron penetration into the liquid helium \cite{Grimes1976a,Stern1978,Rama1988,Cheng1994,Degani2005}. It is important that, as it follows from the previous work and from Eq.~(\ref{eq:asymptotic}), all contributions to the $T=0$ level shift are small.

Understanding the experiment requires finding the $T$-dependent part of the shift of the electron energy levels.
For $T\gtrsim$ 10~mK  the kinematic contribution (\ref{eq:kinematic}) is the dominating part of this shift. From Fig.~\ref{fig:sketch}(d), this shift is small even before the renormalization. However, it is directly observable.  In the approximation (\ref{eq:asymptotic}), the $T$-dependent  part $\Delta E_{n\pb T}^{\rm kin}$ of the kinematic level shift is
\begin{align}
\label{eq:T_dependence}
&\Delta E_{n\pb T}^{\rm kin}\approx A_{n}^{\rm kin}(k_B T)^{5/3}, 
\end{align}
The coefficient $A_{n}^{\rm kin}\approx c^{\rm kin}|\partial_z\psi_{n}|_{z=0}^2 (\hbar^4\rho/\sigma^4)^{1/3}/m$ ($c^{\rm kin}=\Gamma(5/3)\zeta(5/3)\approx 0.1$) sensitively depends on the electron state $n$. Equation (\ref{eq:T_dependence}) predicts a power-law dependence of the level shift on $T$ with exponent 5/3. 
\begin{figure}[h]
\centering
\includegraphics[width=5.5cm]{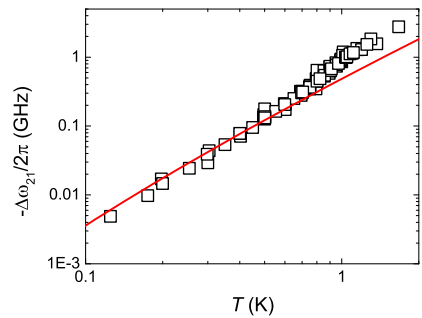}
\caption{\label{fig:3} Temperature dependence of the $1\to  2$ transition frequency calculated for $E_{\perp}=106$~V/cm (solid line). The squares are the experimental results \cite{Collin2017}. In the studied density range $0.67\times 10^7 - 2.4\times 10^7$~cm$^{-2}$ no dependence on the electron density was found, and the data for different densities are combined. } 
\label{fig:comparison}
\end{figure}

The dependence of the energy renormalization on the level number $n$ leads to a temperature-dependent shift of the peaks of microwave absorption due to $n\to n'$ transitions.  Since the characteristic ripplon momenta $\hbar q$ in Eq.~(\ref{eq:kinematic}) largely exceed the thermal electron momentum, the level shift is essentially independent of the electron momentum $\pb$. 

In Fig~\ref{fig:comparison} we present the results for the thermally-induced shift $\Delta \omega_{21}$ of the transition frequency $\omega_{21}=(E_2-E_1)/\hbar$. It is calculated as a sum of the kinematic contribution (\ref{eq:kinematic}) and the contribution from the electrostatic coupling given in SM, keeping only the $T$-dependent terms in the both expressions. To compare the theory with the experiment, the experimentally measured transition frequency \cite{Collin2017} was extrapolated to $T=0$ and the shift was counted off from the extrapolated value. The calculated $T$-dependent frequency shift is free from the ambiguity related to the form of the electron potential at the atomic distance from the helium surface. 

The theoretical curve in Fig.~\ref{fig:comparison} is in excellent agreement with the experimental data shown by squares, with no adjustable parameters. A deviation is observed only  for $T\gtrsim 1$~K, where scattering by helium vapor atoms becomes substantial. The simple expression (\ref{eq:T_dependence}) gives $\Delta \omega_{21}$ that differs from the numerical result by a factor $\sim 3$. 

Two-rippon coupling has been attracting much interest as a mechanism of electron energy relaxation \cite{Monarkha2004,Dykman1978c,Monarkha1978,Dykman2003,Monarkha2010}. Its importance is a consequence of the slowness of ripplons, which makes single-ripplon scattering essentially elastic. In contrast, for two-ripplon scattering, the total wave number of the participating ripplons $|\qb_1 + \qb_2|$ can be of the order of the reciprocal electron thermal wavelength or the reciprocal magnetic length, whereas the wave number of each ripplon $q_{1,2}$ can be much larger,  so that the ripplon energies $\hbar\omega_{q_1}\approx \hbar\omega_{q_2}$ can be comparable to $k_BT$, the intersubband energy gap $|E_n-E_{n'}|$, or the Landau level spacing. 

From Eq.~(\ref{eq:transformed_kinetic}), the matrix element of an electron transition $|n,\pb\rangle \to |n',\pb'\rangle$  calculated for the direct kinematic two-ripplon coupling $\hat H_i^{(2)}$ is $\propto q_1q_2\approx - q_1^2$.  It implies a high rate of deeply inelastic electron relaxation for large $q_{1,2}$. The single-electron kinematic coupling to ripplons $\hat H_i^{(1)}$  very strongly reduces the scattering rate. Using the Bethe trick, one can show that the term $\propto q_1q_2$ drops out from the transition matrix element calculated to the second order in $\hat H_i^{(1)}$. A similar cancellation occurs for the electrostatic electron-ripplon coupling [33]. This strongly reduces the calculated energy relaxation rate, bringing it within the realm of the  experiment. The full analysis of the electron energy relaxation requires also taking into account scattering by phonons in helium \cite{Dykman2003}. This analysis is beyond the scope of the present paper.

In conclusion, we have shown that the system of electrons coupled to the quantum field of capillary waves on the helium surface enables studying a condensed-matter analog of the Lamb shift, which in this case is the shift of the subbands of the quantized motion transverse to the surface. As in the case of the Lamb shift, the matrix elements of the electron coupling to the quantum field are known, and there are terms in the expression for the shift that display an ultraviolet divergence. We have shown that the analysis may not be limited to the conventional intra-subband processes.  We have revealed the diverging inter-subband terms and used the Bethe trick to demonstrate that different diverging terms cancel each other to the leading order, making the overall shift small. The considered system makes it possible to study the dependence of the level shift on temperature. Our theoretical results are in excellent agreement with the experimental observations.

DK was supported by an internal grant from Okinawa Institute of Science and Technology Graduate University;  KK was supported by  JSPS KAKENHI JP24000007; MJL was supported in part by the EU Human Potential Programme under contract HPRN-CT-2000-00157; MID was supported in part by the NSF (Grant no. DMR-1708331).



\begin{widetext}
\newpage

\begin{center}
\large\bf{SUPPLEMENTAL MATERIAL:\\
Ripplonic Lamb Shift for Electrons on Liquid Helium}
\end{center}

\end{widetext}
The Supplemental Material describes the technical details of the calculations carried out in the main text.

\setcounter{equation}{0}

\section{The WKB wave functions}

In this section we describe the analytical approximation for the energies $E_n$ and the eigenfunctions $\psi_{n}(z)$ of highly excited states of motion normal to the helium surface. For energies $\lesssim 1$~eV and for the distances from the surface $z \gtrsim 1 \AA$, the confining potential above the flat helium surface in Eq.~(\ref{eq:Hamiltonian_flat}) has the form 
\begin{equation}
\label{eq:static_potential}
V(z)=-\Lambda z^{-1}+eE_{\perp}z,  \qquad\Lambda = e^2\frac{\epsilon -1}{4(\epsilon+1)}, 
\end{equation}
where $\epsilon$ is the helium dielectric constant, whereas $V(z)\to\infty $ for $z\to - 0$. We note that the effect of the ripplons on the shift of the electron energy levels is accounted for directly, therefore it would be inconsistent to consider the smearing of the helium surface due to the ripplons. However, Eq.~(\ref{eq:static_potential}) has to be modified on the atomic scale, which contributes to the $T=0$ shift of the electron energy levels \cite{Grimes1976a,Stern1978,Rama1988,Cheng1994,Degani2005}, as indicated in the main text.

We change to dimensionless length, energy, and pressing field,
\begin{equation}
 \zeta=z/r_B, \qquad \varepsilon_{n}=E_{n}/R, \qquad F=(eE_{\perp}r_B)/R,
\label{eq:dimensionless_variables}
\end{equation}
($r_B=\hbar^2/\Lambda m$ and $R=\hbar^2/2mr_B^2$ are the localization length of the ground state and the electron binding energy for $E_{\perp}=0$). We will assume that the dimensionless force is small, $F\lesssim 1$; in the experiment discussed in this paper $F\sim 0.1$.

In the variables (\ref{eq:dimensionless_variables}) the Schr\"odinger equation for the motion normal to the surface becomes
\begin{equation}
\left[ -\frac{d^2}{d \zeta^2} -\frac{2}{\zeta} +F\zeta \right] \psi_{n}(\zeta)  = \varepsilon_{n} \psi_{n}(\zeta) 
\label{eq:Schrodinger_dimensionless}
\end{equation}
with the boundary condition $\psi_{n}(0)=0$.

For large energies $\varepsilon_n\gg 1$ and not too small $\zeta$ the solution can be sought in the WKB form
\begin{align}
\label{eq:WKB}
&\psi_{n}(\zeta)=\frac{C_n}{\sqrt{p(\zeta,\vep_n)}} \sin \left[ S(\varepsilon_n)-\int_0^\zeta p(\zeta',\varepsilon_n)d\zeta' + \frac{\pi}{4}\right], \nonumber\\
&S(\varepsilon_n)= \int\nolimits_0^{\zeta_n}p(\zeta',\varepsilon_n)d\zeta',
\end{align}
where $p(\zeta,\varepsilon)=[\varepsilon + 2/\zeta - F\zeta]^{1/2}$ is the scaled classical momentum of motion in the $z$-direction and $C_n$ is the normalization constant, which we set to be a real number. The value of $\zeta_n$ is given by equation $p(\zeta_n,\varepsilon_n)=0$; for large $\varepsilon_n$ we have $\zeta_n\approx F^{-1}\varepsilon_n + 2\varepsilon_n^{-1} \gg 1$.

The WKB approximation breaks down for small $\zeta$, because the confining potential is singular at $\zeta\to 0$. For the WKB to apply, the de Broglie wavelength should be small compared to the distance on which it changes. This means that Eq.~(\ref{eq:WKB}) applies for $\zeta\gg \varepsilon_n^{-3/4}$.  

For small $\zeta$, where $F\zeta \ll \varepsilon + 2/\zeta$, we can disregard the term $F\zeta$ in Eq.~(\ref{eq:Schrodinger_dimensionless}). The solution of this equation then becomes
\begin{equation}
\label{eq:small_distances}
\psi_n(\zeta) =\tilde C_n\zeta e^{-i\varepsilon_{n}^{1/2}\zeta}{} _1\!F_1 \left( i\varepsilon_{n}^{-1/2}+1, 2, 2i\varepsilon_{n}^{1/2}\zeta \right) + {\rm c.c.},
\end{equation}
where ${_1}\!F_1$ is the confluent hypergeometric function and $\tilde C_n$ is a constant. 
We will see that $\tilde C_n$ can be assumed real, in which case the term $\propto {_1}\!F_1$ and its complex conjugate are equal. 

The solutions (\ref{eq:WKB}) and (\ref{eq:small_distances}) should match in the range where $\varepsilon_n^{3/4}\zeta \gg 1$ and at the same time $F\zeta \ll \varepsilon_n + 2/\zeta$. In the considered case of large $\vep_n$, both solutions should apply in the range  $\vep_n/F\gg\zeta \gg \vep_n^{-1/2}$. For $\varepsilon_n^{1/2}\zeta \gg 1$ we have 
\begin{align*}
&_1\!F_1 \left( i\varepsilon_{n}^{-1/2}+1, 2, 2i\varepsilon_{n}^{1/2}\zeta \right)\approx [1/\Gamma(1+i\varepsilon_n^{-1/2})]\nonumber\\
&\times \exp[2i\varepsilon_n^{1/2}\zeta +  (i\varepsilon_n^{-1/2}-1)\log(2i\varepsilon_n^{1/2}\zeta)],
\end{align*}
so that, assuming $\tilde C_n$ to be real, we have in this range from Eq.~(\ref{eq:small_distances})
\begin{align}
\label{eq:asymptot_from_small}
\psi_n(\zeta)&\approx \tilde C_n\vep_n^{-1/2}e^{-\pi/2\vep_n^{1/2}}\nonumber\\
 &\times \sin\left\{\vep_n^{1/2}\zeta +\vep_n^{-1/2}[\ln(2\vep_n^{1/2}\zeta)+\gamma]\right\},
\end{align}
($\gamma$ is the Euler constant), whereas Eq.~(\ref{eq:WKB}) gives
\begin{align}
\label{eq:WKB_asymptotic}
&\psi_n(\zeta)\approx C_n\vep_n^{-1/4}\sin\left\{S(\vep_n)+\frac{\pi}{4} -\vep_n^{1/2}\zeta \right.\nonumber\\
&\left.- \vep_n^{-1/2}[1+\ln(2\vep_n\zeta)]
\right\}
\end{align}

By comparing Eqs.~(\ref{eq:asymptot_from_small}) and (\ref{eq:WKB_asymptotic}), we obtain
\begin{align}
\label{eq:quantization_condition}
S(\vep_n)=\pi\left(n-\frac{1}{4}\right) +\vep_n^{-1/2}(1-\gamma+\ln\vep_n^{1/2})
\end{align}
with integer $n$;  $\tilde C_n=(-1)^{n+1}C_n\vep_n^{1/4}\exp(\pi/2\vep_n^{1/2})$. 

The classical action $S(\vep)$ can be found for small $F/\vep^2$ and $\vep \gg 1$, where the image potential can be considered a perturbation. The image-potential induced correction is nonanalytic in $\vep$,
\begin{equation}
\label{eq:action_approximate}
S(\vep)\approx \frac{2\vep^{3/2}}{3F} +\vep^{-1/2}\log(c_\vep \vep^2/F), \qquad  F/\vep^2\ll1,
\end{equation}
where constant $c_\vep \approx 20$ is estimated by interpolating the numerical value of $S(\vep)$. Equations (\ref{eq:quantization_condition}) and (\ref{eq:action_approximate}) give the energy spectrum of the excited states with logarithmic corrections from the image potential, $\vep_n\approx \vep_n^{(0)} + \vep_n^{(1)}$,
\begin{align}
\label{eq:energy_correction}
&\vep_n^{(0)}=\left[(3\pi F/2)\left( n-1/4\right)\right]^{2/3}, \nonumber\\
&\vep_n^{(1)} = -\frac{F}{\vep_n^{(0)}}\left[\frac{3}{2}\ln\vep_n^{(0)} -1+\gamma + \ln(c_\vep/F)\right].
\end{align}

To the leading order in $\vep_n^{-1}$, the normalization constant $C_n$ can be calculated disregarding the image potential. From Eq.~(\ref{eq:WKB}), $C_n\approx F^{1/2}\vep_n^{-1/4}$. 

For very large $\vep_n$ one can disregard the image potential in Eq.~(\ref{eq:Schrodinger_dimensionless}) and assume that the electrons are in a triangular well. The solution $\psi_n(\zeta)$ can be sought in terms of the Airy functions as Ai$\bigl(F^{1/3}(\zeta - F^{-1}\vep_n)\bigr)$, with $\vep_n$ determined by the condition $\psi_n(0)=0$. This gives asymptotically the same leading term $\vep_n^{(0)}$ in the expression for the energy, $\vep_n$. The correction from the image potential can be calculated as $-2\Bra{\psi_n}\zeta^{-1}\Ket{\psi_n}$; the result is close to Eq.~(\ref{eq:energy_correction}) for $\vep_n^{(1)}$.

In order to compare the results with the experiment \cite{Collin2002,Collin2017}, we performed a detailed numerical study of the energies and the wave functions for the field $E_\perp =106$~V/cm used in the experiment. We calculated the energies $E_{n}$ as well as the matrix elements of the momentum $p_z$  (and of the electrostatic interaction, see below)  for $n\leq 30$ by numerically solving the Schr\"odinger equation (\ref{eq:Schrodinger_dimensionless}). For larger $n$, one can use Eq.~(\ref{eq:small_distances}) to describe the wave functions in the region of small $\zeta \lesssim 1$, which contribute to the overlap integrals $\langle n|p_z|n_1\rangle$ with $n_1\sim 1$. Expression (\ref{eq:small_distances}) for $\psi_n$ was corrected by multiplying it by an extra factor $\exp(iF\zeta^2/4\vep_n^{1/2})$. This factor accounts for the field-induced term in the WKB expression (\ref{eq:WKB}), which was dropped in Eq.~(\ref{eq:WKB_asymptotic}) to match Eq.~(\ref{eq:small_distances}), which refers to the limit $F\zeta^2 \to 0$.
\begin{figure}[h]
\includegraphics[width =5cm]{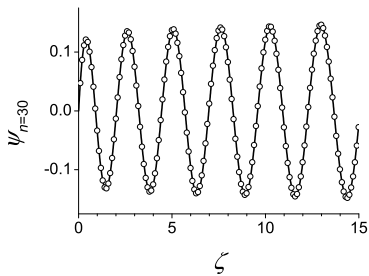}
\caption{Solid line: the analytical expression (\ref{eq:small_distances})  for the wave function of the level $n=30$ at small distance from the surface; the expression has been corrected by the WKB factor $\exp(iF\zeta^2/4\vep_n^{1/2})$. The numerical results are shown by dots. The data refers to $E_\perp = 106$~V/cm, the dimensionless energy of the level is $\varepsilon_{30}=6.57$  }
\label{fig:wave_function_n_30}
\end{figure}

We found that for $n\gtrsim 20$ functions $\psi_n(\zeta)$ modified this way are in a very good agreement with the numerically calculated wave functions in the whole region $\zeta\ll \vep_n/F$, see Fig.\ref{fig:wave_function_n_30}. The values of the matrix elements $\Bra{n}p_z\Ket{n_1}$ calculated with such $\psi_{n}(\zeta)$ for $n_1=1,2$ are in a very good agreement with the numerical values, too, see Table~\ref{table:p}. Therefore in the numerical calculations for large $n$ we used these functions. The matrix elements $\Bra{n}p_z\Ket{n_1}$ become close to those  for $\psi_n$ calculated in the triangular well approximation for $n\gtrsim 10^4$.

\begin{figure}[h]
\includegraphics[width=3.8cm]{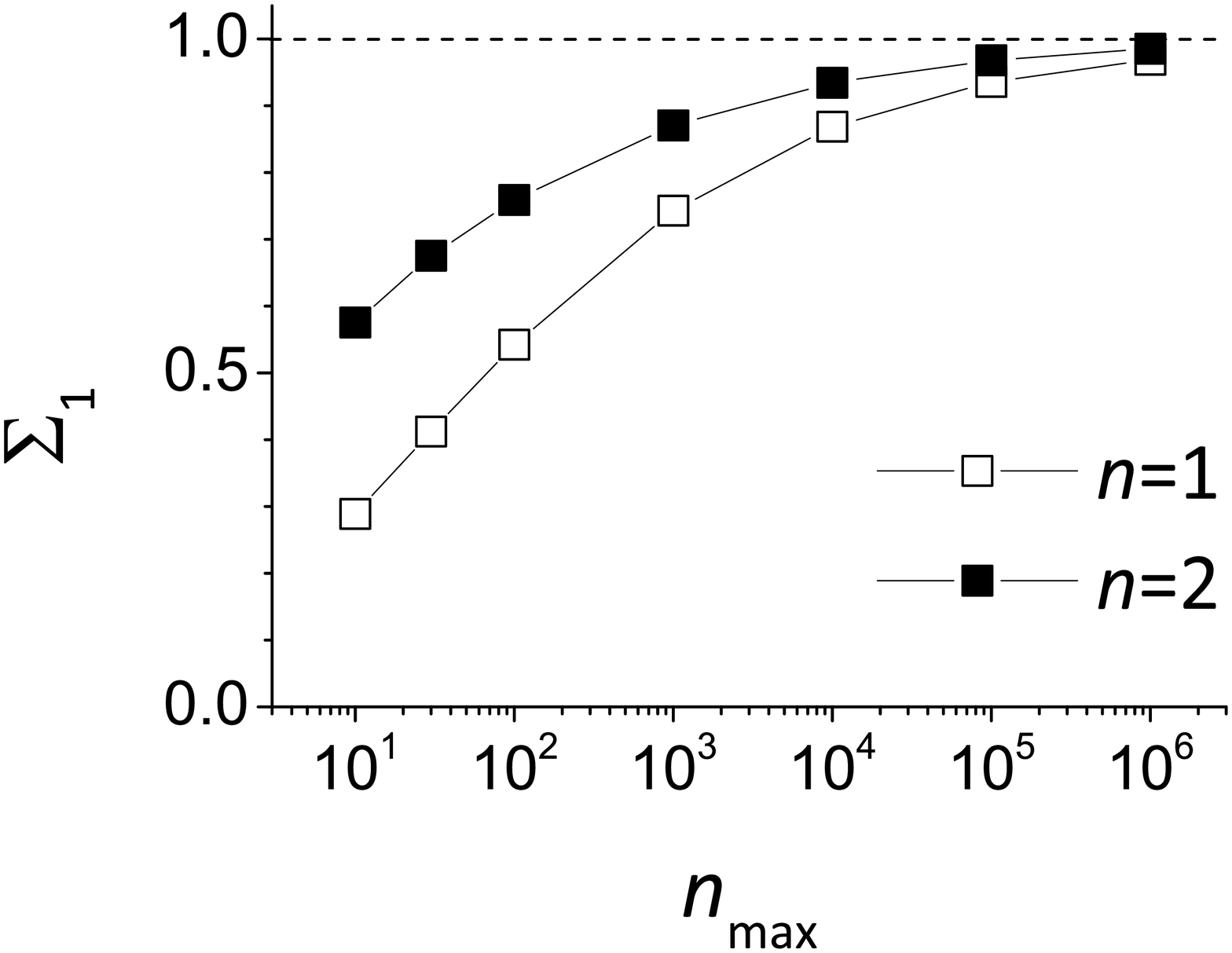}\hfill
\includegraphics[width=3.8cm]{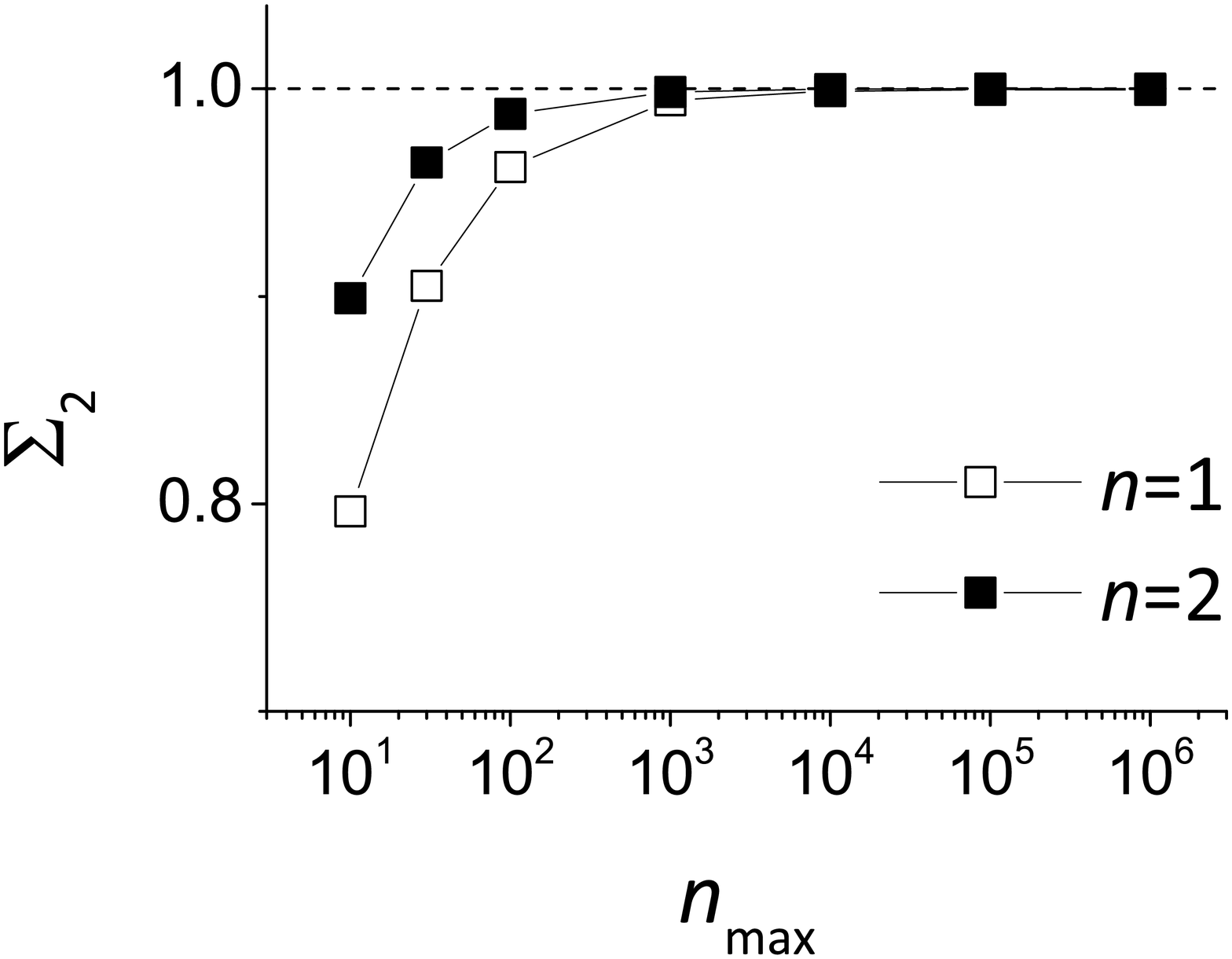}
\caption{The convergence of the numerically evaluated sums (\ref{eq:sum_check}) relevant for finding the energy shift of the states $n=1$ and $n=2$; $\Sigma_1 = \sum_{n'\leq n_{\rm max}}|\langle n'|p_z|n\rangle|^2/ \langle n|p_z^2|n\rangle$ and 
$\Sigma_2 = (2/m)\sum_{n'\leq n_{\rm max}}|\langle n'|p_z|n\rangle|^2/(E_{n'}-E_n)$. The data refer to the  pressing field $E_\perp = 106$~V/cm used in the experiment.}
\label{fig:sum_rules}
\end{figure}

As a test of the accuracy of our numerical calculation we checked the convergence of the sums 
\begin{align}
&\sum_{n'\leq n_{\rm max}}|\langle n'|p_z|n\rangle|^2\to \langle n|p_z^2|n\rangle,  \;n_{\max}\to\infty,
\nonumber\\
&\sum_{n'\leq n_{\rm max},\, E_{n'}\neq E_n}\frac{|\langle n'|p_z|n\rangle|^2}{(E_n-E_{n'})} \to -m/2,\;n_{\max}\to\infty.
\label{eq:sum_check}
\end{align}
The results presented in Fig.~\ref{fig:sum_rules} show slow, but consistent convergence.

\begin{table}
\caption{Values of the scaled matrix elements $\pi_{1n} =(i/\hbar)r_B\langle n|p_z|1\rangle$ and $\pi_{2n} =(i/\hbar) r_B \langle n|p_z|2\rangle$] calculated for $E_\perp =$~106 V/cm and for large $n$ using Eq.~(\ref{eq:small_distances}) with the WKB correcting factor (see Fig.~\ref{fig:wave_function_n_30}) and  the approximation of a triangular potential well. The wave functions $\psi_{1,2}$ are calculated numerically. For comparison, the
numerical solutions of Eq. (\ref{eq:Schrodinger_dimensionless}) for n = 30 give $\pi_{1n}$~= 0.0662 and $\pi_{2n}$~= 0.0415.}
\begin{ruledtabular}
\begin{tabular}{ccc}
$n$ & Equation (\ref{eq:Schrodinger_dimensionless}) & Triangular well \\ [2pt]
& $\pi_{1n}$  \hspace*{14pt} [$\pi_{2n}$] &$\pi_{1n}$ \hspace*{14pt}  [$\pi_{2n}$] \\ [2pt]
30 & 0.0663 \hspace*{8pt} [0.0415] & 0.0740 \hspace*{8pt} [0.0443]\\[2pt]
$10^2$ & 0.0347 \hspace*{8pt} [0.0199] & 0.0419 \hspace*{8pt} [0.0239]\\[2pt]
$10^3$ & 0.0091 \hspace*{8pt} [0.0049] & 0.0104 \hspace*{8pt}[0.0056]\\[2pt]
$10^4$ & 0.0021 \hspace*{8pt} [0.0011] & 0.0023 \hspace*{8pt} [0.0012]\\[2pt]
$10^5$ & 0.0005 \hspace*{8pt} [0.0003] & 0.0005 \hspace*{8pt} [0.0003]\\[2pt]
\end{tabular}
\end{ruledtabular}
\label{table:p}
\end{table}

As expected from the estimate that led to Eq.~(8) of the main text, the matrix elements $\Bra{n}p_z\Ket{n_1}$ fall off approximately as $n^{-2/3}$ for $n\gg 1$ and $n_1\sim 1$. Therefore it was necessary to sum over a large number of the virtual states in Eq.~(7) of the main text. In this equation the summation index is $n'$, and in Eq.~(\ref{eq:sum_check}) and below we  refer to sums over $n'$ for given $n$.  If the sum over $n'$ in Eq.~(7) of the main text is limited to $n' \leq 10^5$, the energy $E_{n'}$ is $\leq 0.98$~eV for the considered $E_\perp = 106$~V/cm, which is on the boundary of the approximation of an infinite potential barrier at the helium surface.

The shape of the electron barrier $V(z)$ on the helium surface for energies $>1$~eV and the form of the electron wave functions for the states with energies $>1$~eV are not known. Also, the shape of the wave function of the low lying states at distances $< 1\AA$ from the surface is not known either. However, we can find the contribution of the highly excited states to the level shift using the sum rules (\ref{eq:sum_check}).

To see how it comes about we note first that, in the $T$-dependent term in Eq.~(7) of the main text for the level shift, the sum over the ripplon wave vectors $\qb$ is limited to $q \lesssim q_T$. Therefore, for temperatures $T\lesssim 0.8$~K, as seen from Fig. 1(b) of the main text, $\Delta_{\pb,\qb,\alpha}< 0.1$~eV. If $E_{n_{\max}}- E_n \gg  \Delta_{\pb,\qb,\alpha}$, in the terms with $n'>n_{\max}$ in Eq.~(7) of the main text one can expand the denominator  in $\Delta_{\pb,\qb,\alpha}/(E_n- E_{n'} )$. The sums over $n'>n_{\max}$ for the terms of the zeroth and first order are then given by the above sum rules with the subtracted terms for $n' < n_{\max}$.
Therefore, once the summation in Eq.~(7) of the main text has been done for $n<n_{\max}$, the overall level shift is known from the sum rules to the first order in the parameter $\delta_{\max} =\Delta_{\pb,q_T,\alpha}/(E_{n_{\max}}- E_n )$  irrespective of the form of the barrier $V(z)$. The second-order term in $\delta_{\max}$ can be easily found also using the sum rule for $|\langle n|p_z|n'\rangle|^2/(E_n-E_{n'})^2$. 

For $E_\perp = 106$~V/cm and $n_{\max}=10^5$, we have $\delta_{\max} \sim 0.1$. Given that the sum rules hold very well for the wave functions we have found, we could therefore extend the summation in Eq.~(7) to $n'\to \infty$ using these wave functions. This is the procedure used to obtain the energy shift in Fig.~2 of the main text.

\section{Contribution of the electrostatic coupling}
\label{sec:electrostatic}

Ripplon-induced warping of the helium surface leads to a change of the electron image potential. The energy of the electron-ripplon coupling is the change of the polarization energy of liquid helium in the electric field of the electron. For the electron located at $\rb$ it has the form \cite{Cole1970,Shikin1974a}
\begin{align}
\label{eq:polarization}
V^{\rm pol}(\rb,z) =& -\frac{\Lambda}{\pi}\int d\rb_1\int_0^{\xi(\rb_1)-\xi(\rb)} dz_1 \nonumber\\
&\times [(\rb - \rb_1)^2 + (z- z_1)^2]^{-2}, 
\end{align}
where $\Lambda = e^2(\epsilon_{\rm He}-1)/8$ ($\epsilon_{\rm He}\approx 1.057$ is the helium dielectric constant and we disregard the terms of higher order in $\epsilon_{\rm He}-1$).

Because the  ratio $\langle \xi^2\rangle^{1/2}/r_B$ is small, one can expand the energy (\ref{eq:polarization}) to the second order in $\xi(\rb)$. One should also take into account another part of the electrostatic energy, which is the energy of the electron in the transverse field $E_\perp$. This energy changes when the electron position is shifted by $\xi(\rb)$. The expression for the total electrostatic part of the electron-ripplon coupling energy then reads
\begin{align}
\label{eq:potential_Hamiltonian}
\hat H_i^{\rm el} &= \sum_\qb V^{(1)}_\qb(z) \xi_\qb e^{i\qb\rb}\nonumber\\
& + \sum_{\qb_1,\qb_2}V_{\qb_1,\qb_2}^{(2)} (z)\xi_{\qb_1}\xi_{\qb_2}e^{i(\qb_1+\qb_2)\rb}.
\end{align}
Here, $\xi_\qb =Q_q(b_\qb + b^\dagger_{-\qb})$, see Eq.~(2) of the main text. Functions $V_\qb^{(1)}$ and $V_{\qb_1\qb_2}^{(2)} $ describe one- and two-ripplon coupling, respectively. They are given in Refs.~\onlinecite{Cole1970,Shikin1974a,Dykman2003},
%
\begin{align}
\label{eq:polariz_coupling}
V_\qb^{(1)}(z)&=\Lambda z^{-2}[1- qz K_1(qz)]+eE_\perp, \nonumber\\
V_{\qb_1,\qb_2}^{(2)}(z)&=-\frac{\Lambda}{z^3}
+\frac{\Lambda}{2z}\left[ q_1^2K_2(q_1z) + q_2^2K_2(q_2z)\right. \nonumber\\
&\left.\qquad- (\qb_1+\qb_2)^2K_2\left(|\qb_1 + \qb_2|z\right)\right]
\end{align}
 $K_{1,2}(x)$ are the modified Bessel functions. In the last term in Eq.~(\ref{eq:polariz_coupling}) for $V^{(2)}_{\qb_1\qb_2}$ one should replace $ (\qb_1+\qb_2)^2K_2\left(|\qb_1 + \qb_2|z\right)$ with $2/z^2$ for $\qb_1 = -\qb_2$. 

The expansion of $V^{\rm pol}(\rb,z)$ in $\xi(\rb)$ breaks down for very small $z$. In this range the major contribution to the integral over $\rb_1$ in Eq.~(\ref{eq:polarization}) comes from small $|\rb_1 - \rb|$. For such $\rb_1$, and assuming that $\xi(\rb)$ is smooth, one can expand $\xi(\rb_1)-\xi(\rb) \approx |\n \xi(\rb)|\cdot |\rb_1-\rb|\cdot\cos\phi$ where $\phi$ is the angle between $\n \xi(\rb)$ and $\rb_1-\rb$. One can then integrate over $|\rb_1-\rb|$ for a given $\phi$ and $z_1$, then over $z_1$, and ultimately over $\phi$. For $z\to 0$, to the leading order in $1/z$ the result reads 
\begin{align}
\label{eq:small_z_nonanalyt}
V^{\rm pol}(\rb,z) \approx &-\frac{\Lambda}{2\pi z}\int_{-\pi/2}^{\pi/2}d\phi \frac{\pi-2\arctan y(\rb,\phi)}{y(\rb,\phi)}, \nonumber \\
&y(\rb,\phi) = [|\n\xi(\rb)|\cos\phi]^{-1}
\end{align}
For small $|\n\xi(\rb)|$, this expression is $\propto [\n \xi(\rb)]^2$, to the leading order in $|\n \xi(\rb)|$. It then coincides with the asymptotic form of $H_i^{\rm el}$ for small $z$, where the major contribution to $H_i^{\rm el}$ comes from $V_{\qb_1,\qb_2} ^{(2)}(z)$. Numerically, approximating Eq.~(\ref{eq:small_z_nonanalyt}) by the term $\propto [\n \xi(\rb)]^2$ works well in a broad range of $|\n\xi(\rb)|$; even for $|\n\xi(\rb)|= 0.7$ the difference with the full expression (\ref{eq:small_z_nonanalyt}) is $<10\%$, and it decreases fast with the decreasing $|\n\xi(\rb)|$. 

The assumption of the smoothness of $\xi(\rb)$ used in Eq.~(\ref{eq:small_z_nonanalyt}) requires that the root mean square displacement $\langle\xi^2(\rb)\rangle^{1/2}$ largely exceed the typical length on which $\xi(\rb)$ changes. If we limit the ripplon wave numbers to $q_c=10^8$~cm$^{-1}$, for the $T=0$ fluctuations we have $q_c\langle\xi^2(\rb)\rangle^{1/2}_{T=0}\sim 1.5$, cf. Fig.~1(c) of the main text. This shows that Eq.~(\ref{eq:small_z_nonanalyt}) is a good approximation for small $ |\n \xi(\rb)|$, whereas for $ |\n \xi(\rb)|\gtrsim 1$ in the expansion of $\xi(\rb_1)-\xi(\rb)$ in Eq.~(\ref{eq:polarization}) one should take into account higher-order terms in $\rb_1 - \rb$. They lead to $V^{\rm pol}(\rb,z)$ increasing even slower than $z^{-1}$ with the decreasing $z$, for very small $z$.



From Eq.~(\ref{eq:small_z_nonanalyt}), the region of small $z$ does not contribute appreciably to the matrix elements of $V^{\rm pol}(\rb, z)$ on the wave functions $\psi_n(z)$, which are $\propto z$ for small $z$.  In the whole region $|\n\xi(\rb)|\leq 1$ the integral of $z^2V^{\rm pol}(\rb,z)$ over the range $z\lesssim \langle \xi^2(\rb)\rangle_{T=0}^{1/2} \sim 10^{-8}$~cm gives an extremely small contribution to the level shift, $\lesssim R\langle \xi^2(\rb)\rangle_{T=0}/r_B^2 \sim 10^{-3}R$ (we keep only the $T=0$ terms in this estimate, since only such terms are essential for large ripplon wave numbers). At the same time, for not too small $z$ the coupling energy (\ref{eq:polarization}) is nonsingular and can be expanded in $\xi(\rb)$. This  justifies using Eqs.~(\ref{eq:potential_Hamiltonian}) and (\ref{eq:polariz_coupling}) to describe the electrostatic electron-ripplon coupling.

The direct two-ripplon coupling $\propto V^{(2)}_{\qb_1,\qb_2}(z)$ leads to a shift of the energy levels already in the first order of the perturbation theory. From Eq.~(\ref{eq:potential_Hamiltonian}),  for an $n$th level the shift is  
%
\[\Delta \tilde E_n^{(2)} = \sum_\qb \Bra{n}V_{\qb,-\qb}^{(2)} (z)\Ket{n}|Q_q|^2(2\bar n_q+1).\]

The terms linear in $\xi_\qb$ give a shift when taken to the second order. This shift is determined by the matrix elements of the sum of the   kinematic interaction $\hat H_i^{(1)}$ [Eq.~(\ref{eq:transformed_kinetic}) of the main text] and the linear in $\xi_\qb$ terms in Eq.~(\ref{eq:potential_Hamiltonian}). To calculate the sum over the intermediate electron states, one can use the Bethe trick, Eq.~(\ref{eq:Bethe}) of the main text. One can then use the relation $i[p_z,V_\qb^{(1)}(z)]=\hbar V_{\qb,-\qb}^{(2)} (z)$, which follows from Eq.~(\ref{eq:polariz_coupling}), to show that the shift $\Delta \tilde E_n^{(2)}$ cancels out (in the calculation one should also use the completeness condition $\sum_{n'}\langle n|\hat A |n'\rangle\langle n'|\hat B|n\rangle = \langle n|\hat A\hat B|n\rangle$ that holds for any operators $\hat A, \hat B$). The full expression for the level shift due to the electrostatic electron-ripplon coupling, which also includes the cross term from the electrostatic and kinematic coupling, is
\begin{widetext}
\begin{align}
\label{eq:level_shift_potential}
\Delta E^{\rm pot}_{n\pb} = \hbar\sum_{\qb,\alpha}{\cal N}_{q\alpha} \sum_{n'}(E_n-E_{n'} - \Delta_{\pb,\qb,\alpha})^{-1}
\left[\hbar \left|\Bra{n'}V^{(1)}_\qb(z)\Ket{n}\right|^2  - 2(E_n-E_{n'}){\rm Im}\,\Bra{n}p_z\Ket{n'} \Bra{n'}V^{(1)}_\qb (z)\Ket{n}\right],
\end{align}
\end{widetext}
where ${\cal N}_{q\alpha}= |Q_q|^2\left[\bar n_q+ (1+\alpha)/2\right]/\hbar^2$, cf. the main text; $\Delta_{\pb,\qb,\alpha}$ is defined in the main text below Eq.~(5). The level shift (\ref{eq:level_shift_potential}) should be added to the purely kinematic shift given by Eq.~(7) of the main text to describe the full level shift due to the electron-ripplon coupling.
Note that the term with $n'=n$ gives the polaronic shift due to virtual processes within the same subband of motion normal to the surface, see \cite{Cole1970,Shikin1971,Jackson1981}. The term quadratic in the pressing field $E_\perp$, which gives the major contribution to this shift for large $E_\perp$ \cite{Jackson1981}, is independent of the subband number $n$ and drops out from the expression for the frequency of inter-subband transitions, which is of the central interest for this paper.


A numerical calculation shows that, for typical pressing fields $E_\perp \lesssim 300$~V/cm, the $T=0$ level shift described by Eq.~(\ref{eq:level_shift_potential}) is much smaller than the $T=0$ kinematic shift discussed in the main text. For $E_\perp = 106$~V/cm (the field used in the experiment discussed in the main text), the $T=0$ term in the sum (\ref{eq:level_shift_potential}) for $n=1$ and $q\gtrsim 10^7$~cm$^{-1}$ is an order of magnitude smaller than the corresponding term in Eq.~(7) of the main text. The $T=0$ shift given by Eq.~(\ref{eq:level_shift_potential}) is a part of the small deviation from the simple model of noninteracting electrons above the flat helium surface. Such deviation has not been measured in the experiment, and it is not easy to measure. This paper shows that the deviation is small without using adjustable parameters or extra approximations.

\section{Two-ripplon scattering}
\label{sec:scattering}

As indicated in the main text, the two-ripplon coupling can provide an important contribution to inelastic electron scattering. This is a consequence of the possibility to scatter into ripplons with large wave numbers $q,q'$ while keeping the total ripplon momentum $\hbar (\qb + \qb')$ small, of the order of the electron thermal momentum or the reciprocal quantum localization length in a magnetic field  multiplied by $\hbar$. The scattering rate is determined by the transition matrix elements calculated for the same total energy and the same total momentum of the electron-ripplon system in the initial and finite states. These matrix elements (the vertex, in terms of the diagrams) are given by the direct two-ripplon coupling in the first order and the single-ripplon coupling in the second order of the perturbation theory. 

If only the direct two-ripplon coupling was kept in the analysis of inelastic scattering, this would lead to a very high scattering rate. For example, the rate of transitions from the bottom of the first excited subband ($n=2, \pb = 0$) to the lowest subband ($n=1$) due to the kinematic coupling $\hat H_i^{(2)}$ [Eq.~(3) of the main text] would be $\gtrsim 10^8$~s$^{-1}$. This is orders of magnitude higher than in the existing experimental data. However, as in the case of the electron energy shift, the major part of the direct coupling is compensated by the linear in $\xi(\rb)$ coupling $\hat H_i^{(1)}$. To find this compensation, one can use again the Bethe trick. 

We consider an electron transition  from the initial state $\Ket{n_i,\pb_i, \{n(\qb)\}}$ to the final state $\Ket{n_f,\pb_f,\{n'(\qb)\}}$, where $|\{n(\qb)\}\rangle$ is the ripplon wave function in the occupation number representation. To the first order in $\hat H_i^{(2)}$ and to the second order in $\hat H_i^{(1)}$, the matrix element ${\cal M}^{\rm kin}_{if}(\qb_1,\qb_2)$ of the kinematic coupling, which describes a two-ripplon transition with emission or absorption of ripplons with wave vectors $\qb_1$ and $\qb_2$, has the form
\begin{widetext}
\begin{align}
\label{eq:matrix_element_kin}
&{\cal M}_{if}^{\rm kin}(\qb_1,\qb_2)= g_{\alpha\beta}\hbar^{-2}Q_{q_1}Q_{q_2}\left[\Bra{n_f}p_z^2\Ket{n_i}(E_{n_i} - E_{n_f})
-\sum_{n'}\Bra{n_f}p_z\Ket{n'}\Bra{n'}p_z\Ket{n_i}(E_{n_i}-E_{n'})\right.\nonumber\\
&\left.\times \left(L_{n_i n'}^{\alpha_1\alpha_2}(\qb_1,\qb_2) + L_{n_i n'}^{\alpha_2\alpha_1}(\qb_2,\qb_1)
\right)\right],\quad 
L_{n_i n'}^{\alpha_1\alpha_2}(\qb_1,\qb_2)= \Delta_{\pb_i +\hbar \qb_1,\qb_2,\alpha_2}/(E_{n_i}-E_{n'} -\Delta_{\pb_i,\qb_1,\alpha_1})
\end{align}
\end{widetext}
Here, the subscripts $\alpha_{1,2}$  indicate whether the transition is accompanied by ripplon emission  ($\alpha_1=\alpha_2=1$), absorption ($\alpha_1=\alpha_2 = -1$), or scattering ($\alpha_1=-\alpha_2=\pm 1$). Factor $g_{\alpha_1\alpha_2} \equiv g_{\alpha_1\alpha_2}(\qb_1,\qb_2)$ is determined by the initial ripplon occupation numbers, $g_{\alpha_1,\alpha_2}(\qb_1,\qb_2)=\{[n(-\alpha_1 \qb_1) +(1+\alpha_1)/2][n(-\alpha_2\qb_2)+(1+\alpha_2)/2]\}^{1/2}$. In the final ripplon state $|\{n'(\qb)\}\rangle$ the occupation numbers of ripplons with the wave vectors $-\alpha_1\qb_1$ and $-\alpha_2\qb_2$ are changed by $\alpha_1$ and $\alpha_2$, respectively, compared to the state $|\{n(\qb)\}\rangle$. In Eq.~(\ref{eq:matrix_element_kin}) we took into account that the final and initial energy of the electron-ripplon system is the same as is also the total in-plane momentum, $\pb_f =\pb_i+\hbar (\qb_1+\qb_2)$.

One can similarly calculate the contribution of the electrostatic electron-ripplon coupling to the matrix element of two-ripplon scattering. An important cancellation of the terms with large ripplon momenta occurs in this case, too. To the first order in the direct electrostatic two-ripplon coupling and to the second order in the one-ripplon electrostatic coupling and the cross-term with the one-ripplon kinematic coupling,  using Eq.~(\ref{eq:polariz_coupling}) we obtain for the matrix element of the same transition as in Eq.~(\ref{eq:matrix_element_kin}) the expression
\begin{widetext}
\begin{align}
\label{eq:matrix_element_pot}
&{\cal M}^{\rm pot}_{if}(\qb_1,\qb2) = g_{\alpha_1\alpha_2} Q_{q_1}Q_{q_2}\left\{\left\langle n_f\left\vert\Lambda\left[\frac{2}{z^3} - \frac{(\qb_1+\qb_2)^2}{z}
K_2(|\qb_1+\qb_2|z)\right]
 - i\left[\frac{E_{n_f}-E_{n_i}}{\hbar\Delta_{\pb_i,\qb_1,\alpha_1}} p_zV^{(1)}_{\qb_1} + (\qb_1 \leftrightharpoons \qb_2)\right]\right\vert n_i\right\rangle\right.\nonumber\\
&\left.+\left[\frac{\langle n_f\vert V^{(1)}_{\qb_2}\vert n'\rangle \langle n'\vert V^{(1)}_{\qb_1} - (i/\hbar) (E_{n_i}- E_{n'} )p_z\vert n_i\rangle}{E_{n_i}-E_{n'}-\Delta_{\pb_i,\qb_1,\alpha_1}}
 -i\frac{(E_{n_i}- E_{n'})\Delta_{\pb_i+\hbar\qb_1,\qb_2,\alpha_2}\Bra{n_f} p_z\Ket{n'} \Bra{n'}V^{(1)}_{\qb_1}\Ket{n_i}}
{\hbar\Delta_{\pb_i,\qb_1,\alpha_1}(E_{n_i}-E_{n'}-\Delta_{\pb_i,\qb_1,\alpha_1})} + (\qb_1 \leftrightharpoons \qb_2)\right] \right\}.
\end{align}
\end{widetext}
Here $V^{(1)}_\qb$ is a shorthand for $V^{(1)}_\qb(z)$ and ``$ + (\qb_1 \leftrightharpoons \qb_2)"$ means adding the same expression with the interchanged $\{\qb_1,\alpha_1\}$ and $\{\qb_2,\alpha_2\}$; function $g_{\alpha_1,\alpha_2}$ is defined below Eq.~(\ref{eq:matrix_element_kin}). Numerical calculations of the relaxation rate based on Eqs.~(\ref{eq:matrix_element_kin}) and (\ref{eq:matrix_element_pot}) are beyond the scope of the present paper.

\bibliographystyle{apsrev4-1}
%

\end{document}